\documentclass[times]{fldauth}

\usepackage{moreverb}
\usepackage{lineno}

\usepackage[colorlinks,bookmarksopen,bookmarksnumbered,citecolor=red,urlcolor=red]{hyperref}

\newcommand\BibTeX{{\rmfamily B\kern-.05em \textsc{i\kern-.025em b}\kern-.08em
T\kern-.1667em\lower.7ex\hbox{E}\kern-.125emX}}

\def\volumeyear{2010}

\newcommand\Rey{\mbox{\textit{Re}}}  

\begin{document}

\runningheads{N.~D.~Sandham, R.~Johnstone, C.~T.~Jacobs}{Surface-sampled simulations of turbulent flow}

\title{Surface-sampled simulations of turbulent flow at high Reynolds number}

\author{Neil~D.~Sandham, Roderick~Johnstone, Christian~T.~Jacobs\corrauth}

\address{Faculty of Engineering and the Environment, University of Southampton, University Road, Southampton, SO17 1BJ, United Kingdom}

\corraddr{Faculty of Engineering and the Environment, University of Southampton, University Road, Southampton, SO17 1BJ, United Kingdom.}

\begin{abstract}
A new approach to turbulence simulation, based on a combination of large-eddy simulation (LES) for the whole flow and an array of non-space-filling quasi-direct numerical simulations (QDNS), which sample the response of near-wall turbulence to large-scale forcing, is proposed and evaluated. The technique overcomes some of the cost limitations of turbulence simulation, since the main flow is treated with a coarse-grid LES, with the equivalent of wall functions supplied by the near-wall sampled QDNS. Two cases are tested, at friction Reynolds number $\Rey_\tau=4200$ and $20\,000$. The total grid node count for the first case is less than half a million and less than two million for the second case, with the calculations only requiring a desktop computer. A good agreement with published DNS is found at $\Rey_\tau=4200$, both in terms of the mean velocity profile and the streamwise velocity fluctuation statistics, which correctly show a substantial increase in near-wall turbulence levels due to a modulation of near-wall streaks by large-scale structures. The trend continues at $\Rey_\tau=20\,000$, in agreement with experiment, which represents one of the major achievements of the new approach. A number of detailed aspects of the model, including numerical resolution, LES-QDNS coupling strategy and sub-grid model are explored. A low level of grid sensitivity is demonstrated for both the QDNS and LES aspects. Since the method does not assume a law of the wall, it can in principle be applied to flows that are out of equilibrium.
\end{abstract}


\maketitle

\vspace{-6pt}

\section{Introduction}
Despite advances in hardware and in particular the use of massively parallel supercomputers, applications of direct numerical simulation (DNS) are limited in terms of the Reynolds number ($\Rey$) that can be reached, owing to the cost of the simulations. Measured in terms of number of grid points, the cost scales strongly with $\Rey$, for example the number of grid points required scales as $\Rey_L^{37/14}$ (where $L$ is the distance from the leading edge) for boundary layer flow \cite{ChoiMoin_2012} and smaller timesteps are also required as the grid becomes finer. A cheaper approach is large-eddy simulation (LES) where only the larger scales are simulated, while smaller scales are modelled. However, near a wall the smaller scales play a predominant role and to obtain sufficient accuracy many LES in practice end up being `wall-resolved' LES, where grid node counts are significantly lower than DNS (typically of the order of 1\%) but a strong scaling with $\Rey$ remains, meaning that LES is also too expensive for routine application, for example to flow over a commercial aircraft wing. The alternative of wall-modelled LES has much more attractive scaling characteristics (fixed in terms of boundary layer thickness, for example), but relies very heavily on a wall treatment. Given that there is no accurate reduced-order model for turbulence near a wall (which would require some kind of breakthrough solution of the `turbulence problem'), a lot of reliance would be placed on the near-wall model, with little likelihood of significant improvements over second-moment closure approaches based on the Reynolds-averaged equations. In this paper we consider an alternative approach whereby small-domain simulations are used to represent the near-wall turbulence, in a non-space-filling manner, and linked to an LES away from the wall, where the sub-grid models might be expected to work with reasonable accuracy. 

To understand the new approach, an appreciation of recent progress in understanding the physics of near-wall turbulence is useful. The inner region, consisting of the viscous sublayer and the buffer layer, out to a wall-normal distance of $z^+\approx 100$ (where $z$ is the wall normal distance and the dimensionless form is $z^+=z u_\tau/\nu$, where $\nu$ is the kinematic viscosity and $u_\tau=\sqrt{\tau_w/\rho}$ is the friction velocity, with $\tau_w=\mu \left( du/dz \right)_w$ the wall shear stress, $\mu=\rho \nu$ being the viscosity and $\rho$ the density) follows a known regeneration cycle \cite{JimenezPinelli1999}, whereby vortices develop streamwise streaks, which give rise to instabilities that create new vortices. The streamwise scales are up to 1000 in wall units ($\nu/u_\tau$), while the spanwise scale is 100 (sufficient to sustain near-wall turbulent cycles \cite{JimenezMoin_1991}), but one should note that the probability distributions are smooth over a range of scales, and the regeneration process doesn't involve single Fourier modes with these wavelengths. The outer region of a turbulent flow follows a different known scaling, where a defect velocity (relative to the centreline in internal flows, or the external velocity in boundary layers) scales with $u_\tau$ and the geometry of the flow (for example boundary layer thickness). As the Reynolds number is increased an overlap between these inner and outer-layers is found and, at very high $\textrm{Re}$, recent pipe flow experiments \cite{Hultmark2012} provide good evidence for a logarithmic region in the mean velocity profile. 

Within the logarithmic region of turbulent boundary layers, pipes and channels very large scale motions (VLSMs) (sometimes referred to as `superstructures') have been observed, for example in \cite{Monty2007}. These structures are in addition to the near-wall turbulence cycle and possible organised motions in the outer part of the flow. Interestingly these VLSM structures are longer than those of the outer layer \cite{Adrian_etal_2000, Adrian_2007, Monty2007, Moarref2013}. The presence of both outer-layer motions and VLSMs means that the near-wall flow cannot be considered as a separate feature, but one that is modulated by larger-scale flow features. This leads to increases in the near-wall fluctuations as $\Rey$ is increased, as has been shown experimentally. For example \cite{Hultmark2012} shows a small increase in the near wall ($z^+=12$) peak in streamwise fluctuation level and a much larger increase for $z^+>100$, eventually leading to a separate peak in the fluctuation profile.

Further insight into the near-wall structure of turbulent flow has been obtained recently from a resolvent-mode analysis of the mean flow \cite{McKeonSharma2010}. The resolvent modes are obtained from a singular value decomposition of the linearised Navier-Stokes equations subject to forcing and shows the response of the flow. From this type of analysis, Moarref et al. \cite{Moarref2013} extracted near-wall, outer layer and mixed scalings. In particular, at very high Reynolds number three kinds of structures were shown to be present, including a near-wall structure whose scaling was in good agreement with the regeneration cycle discussed above. In the outer region the spanwise width of structures was shown to scale with the channel half height, whereas in the logarithmic region the width had a mixed scaling. Given these insights into the key structures in turbulent wall-bounded flow, it is interesting to consider a simulation approach based on resolving these classes of structures.

There have been a small number of previous attempts to combine different simulations to resolve the various layers of flow near a wall. A multi-block approach was developed by Pascarelli et al. \cite{Pascarelli2000}. This method includes a multi-layer structure with a large block covering the channel central region and smaller blocks near the wall that were periodically-replicated. Simulations were only carried out at low $\Rey$ but it was observed that the flow adjusted very quickly to the imposition of periodic spanwise boundary conditions at the block interfaces. The method envisioned more layers at high $\Rey$. The cost saving at the $\Rey$ simulated was found to be modest and the method would not capture the modulation of small scales by large scales, since the same near-wall box was used everywhere. Another approach has been proposed recently \cite{Tang2013} in which a minimal flow unit for near-wall turbulence is coupled to a coarse-grid LES for the whole domain, with a rescaling of both simulation at each timestep. It is not clear from the description whether the minimal flow simulation feeds back the correct local shear stresses to the large structures, but results from this approach are shown to reproduce experimental correlations for skin friction \cite{Dean1978} within 5\% up to $\Rey_\tau=10,000$.

\begin{figure}[t]
  \centering
  \includegraphics[width=0.7\textwidth]{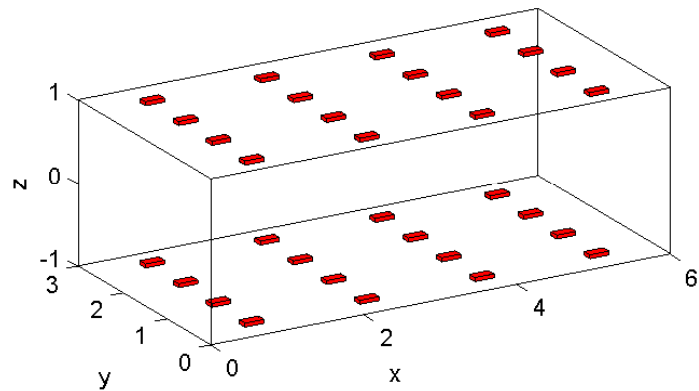}
  \caption{Schematic of the computational arrangement for simulation of turbulent channel flow. The outer box is the LES domain, while the red-shaded boxes are the computational domains for the quasi-DNS.}
  \label{fig:channel}
\end{figure}

In the present contribution we consider an approach that uses multiple near-wall simulations that are able locally to respond to changes in the outer-layer environment, provided by an LES. In return the near-wall simulations provide the wall shear stress required by the LES as a boundary condition. The general arrangement is sketched on Figure \ref{fig:channel} for a simulation of turbulent channel flow. In effect the set of near-wall simulations (shown in red on the figure) are used as the near-wall model. However these simulations are only sampled (not continuous) in space, hence a large saving in computational cost is possible. As a shorthand notation we will refer to the near-wall simulations as quasi-DNS (QDNS) since no sub-grid model is used, but resolutions do not need to be fine enough for these to be fully-resolved DNS. The approach proposed here follows the style of heterogeneous multiscale methods (HMM), a general framework in which different modelling techniques/algorithms are applied to different scales and/or areas of the computational grid \cite{EEngquist_2003, E_etal_2003, E_etal_2007, LeeEngquist_2016}. More specifically, the crux of HMM is the coupling of an overall macroscale model (i.e. the LES in this case) with several microscale models (i.e. the QDNS blocks); it is these microscale models that can provide missing/more accurate data (i.e. the shear stress boundary conditions) back to the macroscale model.

A similar multiscale reduced-order approach was formulated independently by \cite{Grooms_etal_2015} and applied to a quasigeostrophic model of the Antarctic Cirumpolar Current. The velocity field and potential vorticity gradient were advanced in time using a coarse grid model; this model comprised small embedded subdomains at each `coarse' grid location (in contrast to the use of blocks encompassing multiple coarse grid points in this work), within which smaller-scale eddies evolved on a separate spatial and temporal scale. This is similar to the method proposed here in that the domain comprises smaller turbulence-resolving simulations that are coupled with a coarser grid simulation. Furthermore, the components of the eddy potential vorticity flux divergence were computed and averaged in the subdomains and fed back to the coarse grid model, much like the near-wall averaged shear stresses computed in the approach described here. However, unlike the present work, the state of the eddy-resolving embedded subdomains was not carried over between coarse grid time-steps and was reset each time to a given initial condition.

In this paper we set out the method and present results from a proof-of-concept simulation of turbulent channel flow, also showing the sensitivity of the method to various numerical parameters. Section \ref{sect:numerical_formulation} provides details of the numerical approach and its implementation as a Fortran code. Section \ref{sect:sensitivity} presents the proof-of-concept results from the simulation of turbulent channel flow. The potential for extension of the method to very high Reynolds number is then discussed in Section \ref{sect:higher_Re}. The paper closes with some conclusions in Section \ref{sect:conclusions}.

\section{Numerical formulation}\label{sect:numerical_formulation}

\subsection{Numerical method}
The same numerical method is used for both the LES and the near-wall QDNS domains shown in Figure \ref{fig:channel}, all of which have periodic boundary conditions applied in the wall-parallel directions $x$ and $y$. Within these domains the incompressible Navier-Stokes equations are solved on stretched (in $z$) grids, using staggered variables (with pressure $p$ defined at the cell centre and velocity components $u_i$ at the centres of the faces), by an Adams-Bashforth method. The governing equations are the continuity equation

\begin{equation}
	\frac{\partial u_i}{\partial x_i}=0
\end{equation}
and the momentum equations

\begin{equation}
 \frac{\partial u_i}{\partial t}+	\frac{\partial u_i u_j}{\partial x_j}=\delta_{i1}-\frac{\partial p}{\partial x_i} + \frac{1}{\Rey_\tau} \frac{\partial^2 u_i}{\partial x_j \partial x_j},
\end{equation}
where all variable are dimensionless (normalised using the channel half height, friction velocity, density and kinematic viscosity) and the term $\delta_{i1}$ provides the driving pressure gradient. Enforcing a constant pressure gradient or constant mass flow rate are the two main approaches to ensuring that the flow field evolves with a near-constant wall shear velocity \cite{Berselli_etal_2006}. In the present work, the constant pressure gradient $\delta_{i1}$ frequently used in similar channel flow simulation setups (e.g. \cite{Kim_etal_1987, OrlandiLeonardi_2008, Busse_etal_2015}) is used only in the LES, whereas for the QDNSs it is set to zero and a constant mass flow rate is employed for consistency reasons so that there is conservation of mass between the LES and QDNS. It was found that using only the LES stresses alone to drive the QDNS simulations resulted in too high a flow velocity. Any inaccuracies in the shear stresses would increase over time since there was no mechanism in place to keep the wall shear velocity (and therefore $Re_\tau$) near the desired constant value.

Grids are uniform in the wall-parallel directions $x$ and $y$ and stretched in the wall-normal ($z$) direction according to

\begin{equation}
z=\frac{\tanh(a\zeta)}{\tanh(a)},
\end{equation}
where $a$ is a stretching parameter and $\zeta$ is uniformly spaced on an appropriate interval ($-1 \leq \zeta \leq1$ in the LES for example).

The Adams-Bashforth method advances the solution in time using two steps. In the first step a provisional update of the velocity field is made according to 

\begin{equation}
	u_i^*=u_i^n+\Delta t \left[ \frac{3}{2}H_i^n -  \frac{1}{2}H_i^{n-1}+\frac{1}{2}\frac{\partial p^{n-1}}{\partial x_i} +\delta_{i1} \right],
\end{equation}
where 

\begin{equation}
	H_i=-\frac{\partial u_i u_j}{\partial x_j}+\frac{1}{\textrm{Re}} \frac{\partial^2 u_i}{\partial x_j \partial x_j}.
\end{equation}
A final correction is then made to give

\begin{equation}
	u_i^{n+1}=u_i^*-\frac{3}{2}\Delta t \frac {\partial p^n}{\partial x_i},
\end{equation}
where the pressure is obtained by solution of

\begin{equation}
	\frac{\partial^2 p}{\partial x_i \partial x_i}=\frac{2}{3\Delta t} \frac{\partial u_i^*}{\partial x_i}.
\end{equation}
Application of a fast Fourier transform in horizontal planes leads to a tridiagonal matrix that is solved directly.

\subsection{Model implementation}
The model code was written in Fortran 90, with conditional statements used to enable/disable the LES parameterisation depending on the flag set in the simulation setup/configuration file. Each iteration of the combined LES-QDNS approach entailed first running each QDNS simulation individually with its own setup file (containing the number of timesteps to perform, for example); the LES was then run immediately afterwards to complete the iteration (and thus a single LES timestep, as explained in the next subsection). The setup and execution of these simulations was performed using a Python script that ensured the simulations were run in the correct order, and also performed statistical averaging and postprocessing of the simulation results. Such postprocessing includes the averaging of the shear stresses from all the QDNS and writing out these results to a file in a format that the LES expects, as discussed in the next section. Note that, while the model itself was written in Fortran and could only be executed in serial, the Python script that handled the execution of the simulations was parallelised such that all of the QDNS were executed at the same time, with the results then being combined/postprocessed via MPI Send/Receive operations. The mpi4py library \cite{Dalcin_etal_2005} was used for this purpose. For a setup involving $N \times N$ QDNS per wall, the LES-QDNS approach requires $(N \times N \times 2) + 1$ MPI processes ($N \times N \times 2$ processes for the total number of QDNS, and one process for the LES).

\subsection{Interconnection between LES and QDNS}
The basic arrangement for the simulations is as shown on figure \ref{fig:channel}. To illustrate the details we consider a baseline case at $\Rey_\tau=4200$, corresponding to the highest current $\Rey_\tau$ for DNS of channel flow \cite{LozanoDuran2014}. The DNS used a domain of size $2\pi$ by $\pi$ by 2 with a $2048 \times 2048 \times 1081$ grid. The smallest resolved length scale in a DNS needs to be $O(\eta)$, where $\eta$ is the Kolmogorov length scale \cite{MoinMahesh_1998}. The choice of $O(\eta)$ grid spacing in the DNS of \cite{LozanoDuran2014} therefore satisfied this requirement, and is consistent with known guidelines for the choice of wall units in turbulent channel flow simulations (see e.g. \cite{Kim_etal_1987, Moser_etal_1999, LeeMoser_2015}). Here we attempt the same configuration using an LES in a domain $6 \times 3 \times 2$\footnote{Note that the domain size of $6 \times 3 \times 2$ did not match exactly with the DNS domain size of $2\pi \times \pi \times 2$ because such round numbers were convenient for wall unit measurements and choice of QDNS block size. The results were found not to be sensitive to this small inconsistency.} on a $24\times 24 \times 42$ grid (with stretching parameter $a$ set to 1.577) with a $4\times 4$ array of QDNS on each wall, each QDNS using a $24^3$ grid (with stretching parameter $a$ set to 1.4) covering a domain in wall units of $1000 \times 500 \times 200$. The total number of grid points is less than half a million, or 0.01\% of the DNS. In this baseline case the QDNS grid spacing in wall units is $\Delta x^+=41.7$, $\Delta y^+=20$ with the first cell centre at $z^+=1.5$. 

The choice of QDNS resolution follows guideline values in the literature (e.g. $\Delta x^+$ typically less than 50 in the spanwise direction compared to 20 for DNS \cite{Spalart_2000}) such that the cost of the QDNS is approximately an order of magnitude less than a full DNS near the wall \cite{Sagaut_etal_2013}. It was found that refining this further had little impact on the accuracy of the results, as discussed in Section \ref{sect:sensitivity}. Seen in plan view the entire QDNS occupies one LES cell (i.e.~$L_{x,\textrm{QDNS}}= \Delta x_\textrm{LES}$ and $L_{y,\textrm{QDNS}} = \Delta y_\textrm{LES}$. In the wall-normal direction the QDNS overlaps the LES, in this case by three cells, to avoid using the immediate near-wall points that are most susceptible to errors in the accuracy of the sub-grid modelling. These three cells cover the region out to $z^+=200$ with the centre of the first LES cell at $z^+=30$. The LES grid was deliberately kept very coarse in order to highlight the potential savings of the proposed method and how it takes advantage of the separation of scales, although it was found \textit{a posteriori} that it needed refining to a $96 \times 96 \times 56$ grid in order to yield a much better mean flow prediction (see Section \ref{sect:sensitivity}).

The required resolution for DNS and QDNS scales strongly with Reynolds number \cite{Spalart_2000}, with the number of DNS grid points being proportional to $\Rey^{9/4}$ \cite{RogalloMoin_1984} (or $\Rey_L^{37/14}$ in the more recent calculations of \cite{ChoiMoin_2012}). The resolution requirements for QDNS are likely to be similar to that of wall-resolving LES which scales proportional to $\sim \Rey^2$ \cite{Chapman_1979, Leschziner_etal_2009, ChoiMoin_2012}, while wall-modelled LES scales weakly with Reynolds number ($\Rey^{2/5}$ \cite{Chapman_1979, ChoiMoin_2012}). In terms of resolving the turbulence structures, small-scale eddies and streaks near the wall scale with wall units while the LSMs scale with domain size \cite{Moarref2013}.

The time step for the QDNS is set to $\Delta t=0.0001$ and 25 QDNS steps are run before one LES update (i.e.~the LES operates on a timestep of 0.0025). The respective Courant number criteria need to be respected for both the LES and QDNS simulations, which determines the number of QDNS steps per LES step.

\begin{figure}[t]
  \centering
  \includegraphics[width=0.7\textwidth]{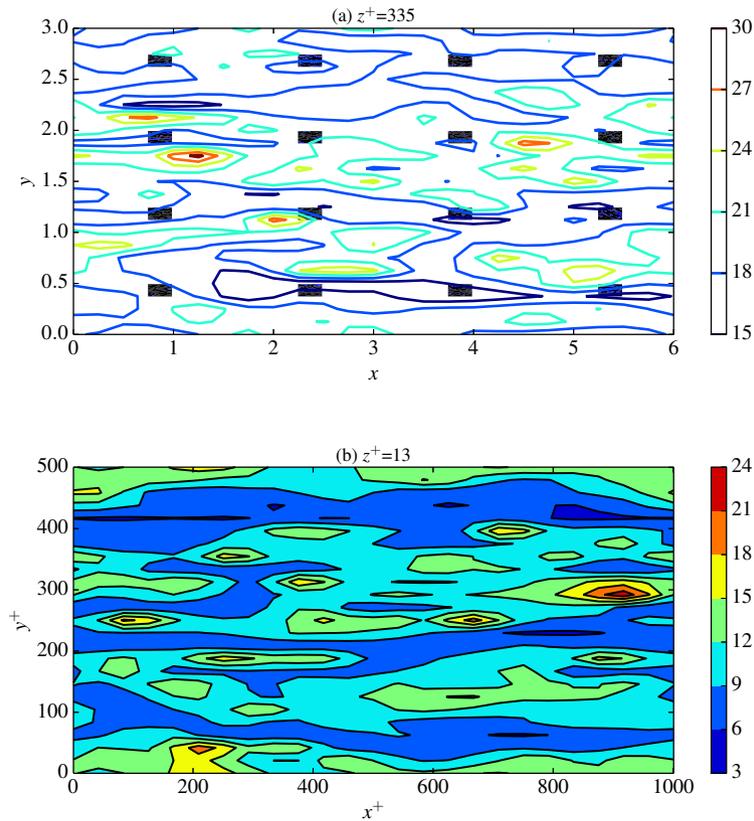}
  \caption{Plan view showing (a) streamwise velocity contour lines at $z^+=335$ from the LES at $\textrm{Re}_\tau=4200$ with the dark areas showing the locations of the QDNS domains, (b) expanded view of filled contours of streamwise velocity at $z^+=13$ in one of the QDNS sub-domains.}
  \label{fig:uplus_contours_Re4200}
\end{figure}

The QDNS are driven by the LES. The QDNS are run in constant mass-flux mode with the mass fluxes in $x$ and $y$ provided by the LES. At the upper boundary conditions the QDNS use $w=0$ and apply a viscous stress corresponding to the shear stresses from the LES. This effectively sets $du/dz$ and $dv/dz$ at the upper boundary of the QDNS and, together with the enforced mass flux, drives the QDNS to match the LES in these aspects. Each QDNS is thus driven by the local LES conditions and simulates the response of wall turbulence to large-scales present in the LES. Figure \ref{fig:uplus_contours_Re4200} shows a snapshot of the results from a simulation. The streamwise velocity is shown in a plan view. Part (a) of the figure shows the whole LES domain at $z^+=335$, with QDNS sub-domains visible as the dark areas. Part (b) of the figure zooms in on one of the QDNS domains, showing streamwise velocity contours near the wall ($z^+=13$). In this arrangement it can be seen how the $4 \times 4$ array of QDNS samples the large-scale structures from the LES. At the end of the 25 QDNS time steps the shear stresses $(du/dz)_w$ and $(dv/dz)_w$ are averaged over each QDNS and linearly interpolated back to the LES to provide the lower boundary condition. Such a boundary condition is considered a good first approximation, despite the QDNS blocks not resolving turbulence structures down to the Kolmagorov length scale, because the QDNSs are capable of resolving the near-wall streaks to reduce the empiricism required at the wall \cite{Spalart_etal_1997, Spalart_2000}. It may be more desirable to use more information from the QDNS (e.g. transferring all components of the Reynolds stress tensor back to the LES and computing a contribution to the eddy viscosity for use in the LES) to obtain a more accurate result. Nevertheless, the current sampling technique and the interpolation back to the full LES domain is advantageous since it exploits the emerging spectral gap that exists between the large and small scales at large Reynolds number \cite{Moarref2013}.

Larger domains are handled by increasing the size of the LES domain and increasing the number of QDNS blocks. It should be noted that there is only a very small amount of communication between the LES and QDNS calculations (four floating point numbers into each QDNS and two returned per 25 steps of computational effort). Thus the introduction of the QDNS subdomains brings with it an additional level of parallelism, with parallel treatment also possible within the LES and QDNS blocks using conventional strategies.

Once fully developed, the turbulent dynamics are homogeneous in the spanwise and streamwise directions \cite{Kim_etal_1987} and thus the use of a regular grid on each wall is a justifiable initial choice. However, instead of keeping the QDNS blocks stationary, it may be more appropriate to move the blocks downstream with the flow speed in an attempt to track smaller-scale turbulent structures. It is possible that the effects of these turbulent small-scale structures are being dissipated by the averaging procedure or simply by the region of lower resolution outside the QDNS block, with downstream blocks becoming increasingly inaccurate as a result. It is unclear how many QDNS blocks will be required in general, but the number is likely to scale with $\Rey_\tau$ in order to obtain adequate sampling near the wall.

\section{Proof of concept and sensitivity to numerical parameters}\label{sect:sensitivity}
\begin{figure}[t]
  \centering
  \includegraphics[width=1.0\textwidth]{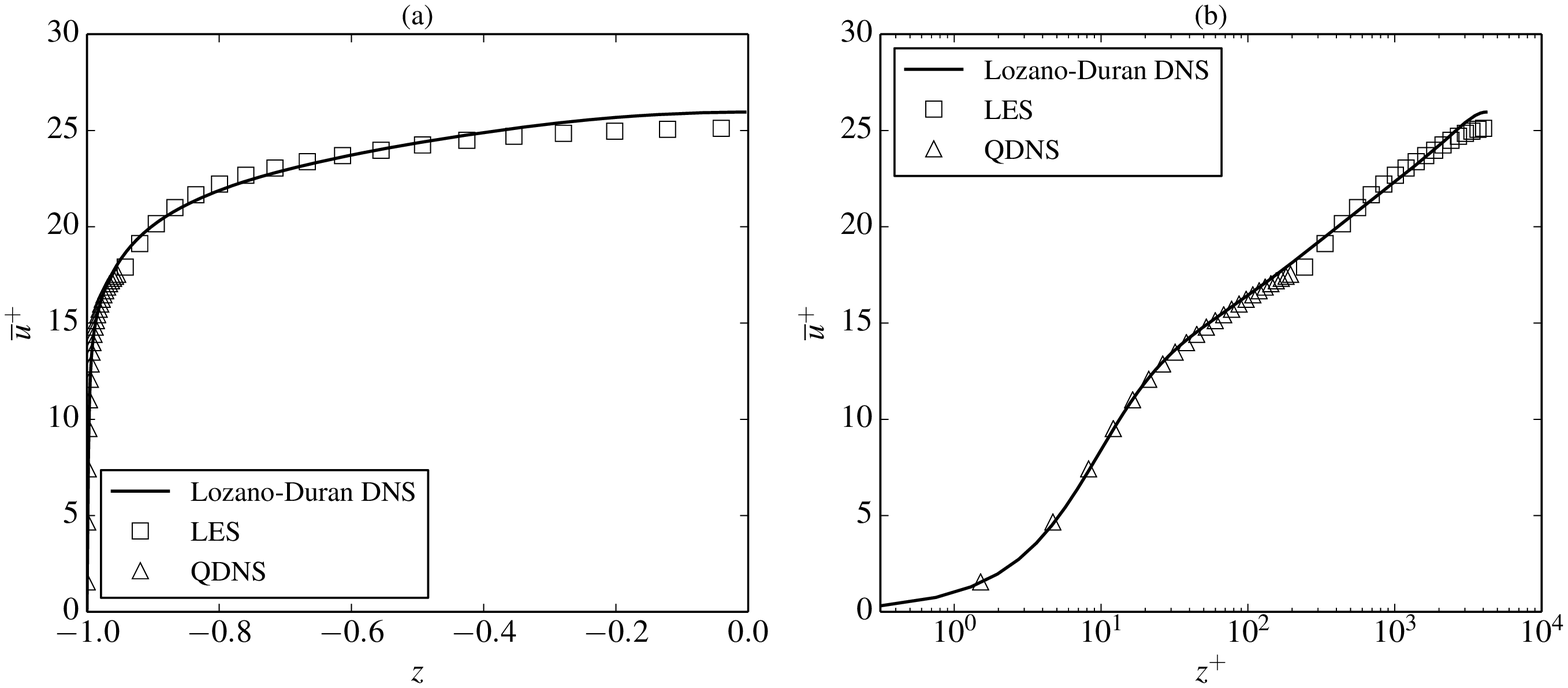}
  \caption{Comparison of the combined LES/QDNS results for mean streamwise velocity with DNS \cite{LozanoDuran2014} (solid line) at $\Rey_\tau=4200$ (a) in linear scale, and (b) in semi-logarithmic co-ordinates. Open triangles show the QDNS (carried out on $24^3$ grids), squares the LES (on a $24\times24\times 42$ grid).}
  \label{fig:mean_uplus_combined}
\end{figure}

\begin{figure}[t]
  \centering
  \includegraphics[width=0.6\textwidth]{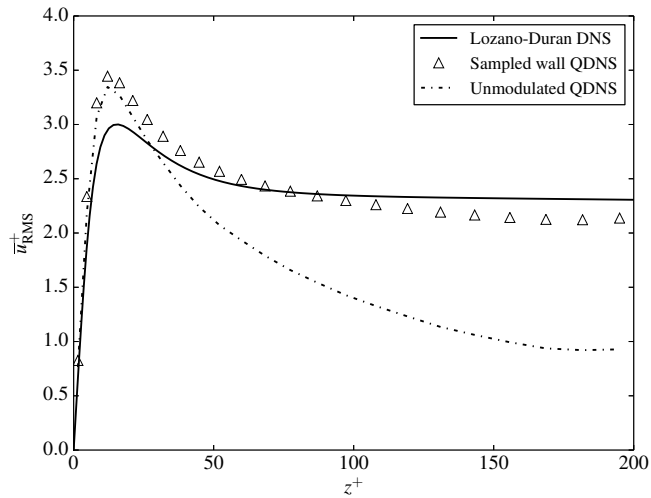}
  \caption{Root-mean-square streamwise velocity in the near-wall region at $\textrm{Re}_\tau=4200$, comparing the QDNS (triangles) from the mixed QDNS-LES simulation with DNS (solid line) and with a separate QDNS, in which the near-wall region is not modulated by structures from the outer region.}
  \label{fig:rms_uplus}
\end{figure}

The mean streamwise velocity $\overline{u}^+$ and root mean square (RMS) of the streamwise velocity fluctuations $\overline{u}^+_\mathrm{RMS}$ were used as performance measures. These are defined, for each point $k$ in the $z$-direction, by

\begin{equation}
   \overline{u}^+ = \frac{1}{SN_xN_y}\sum_{s=1}^{S}\sum_{i=1}^{N_x}\sum_{j=1}^{N_y}u^+_{i,j,k},
\end{equation}
and

\begin{equation}
   \overline{u}^+_\mathrm{RMS} = \sqrt{\left(\frac{1}{SN_xN_y}\sum_{s=1}^{S}\sum_{i=1}^{N_x}\sum_{j=1}^{N_y}{u^+_{i,j,k}}^2\right) - {\overline{u}^{+}}^2}.
\end{equation}
where $u^+_{i,j,k}$ is the dimensionless velocity at grid point $(i,j,k)$. The quantities $N_x$ and $N_y$ are the number of grid points in the $x$ and $y$ directions. The quantities were not accumulated over all time-steps, but were instead accumulated every $S$ timesteps, where $S$ was chosen to be sufficiently small to ensure a steady average. In addition, the mean velocity relative to the friction velocity was also considered. This quantity is defined as

\begin{equation}
   \widetilde{\overline{u}^+} = \frac{1}{2}\int_{-1}^1 \overline{u}^+ \ \mathrm{d}z.
\end{equation}

The mean streamwise velocity for the baseline case is shown in figure \ref{fig:mean_uplus_combined} in linear and semi-logarithmic co-ordinates in parts (a) and (b) respectively, showing a composite of the LES results (with squares, omitting the first 3 cells) and the near-wall QDNS (shown with triangles). Overall a reasonable match to the reference DNS is observed despite the very low grid node count. The QDNS simulations correctly capture the viscous sublayer and buffer layer, while the LES captures the outer layer. Both the QDNS and LES undershoot the reference DNS by about 5\% at the LES/QDNS interface and the LES gives noticeably too low a centreline velocity (by 3\%). The mean velocity relative to the friction velocity is 23.3 which is $\sim$0.9\% lower than the DNS and 2.9\% lower than Dean's correlation \cite{Dean1978}, which together provide a useful measure of the overall accuracy of this approach. With all the data available from the QDNS, it would in principle be possible to improve the near-wall sub-grid modelling in the LES to address the undershoot at the interface (for example the eddy viscosity can be computed from the QDNS and used in the LES), however in the present contribution we use the same (Smagorinsky) sub-grid model for all cases.

An interesting feature emerges when one considers the root mean square (RMS) of streamwise velocity fluctuations from the QDNS simulations, shown on figure \ref{fig:rms_uplus}. To assemble this figure, as with the QDNS shown in figure \ref{fig:mean_uplus_combined}, all 16 QDNS on one wall were averaged in horizontal planes and over time. The result is generally in good agreement with the DNS. There is an overshoot in the peak at $z^+=12$, which is likely due to under-resolution within the QDNS blocks; similar over-shoots have been observed in the RMS streamwise velocity for large eddy simulations of turbulent channel flow (at lower $\Rey_\tau$ values of 180, 395 and 640) where the near-wall zone is not adequately resolved by the grid \cite{Veloudis_etal_2008, Singh_etal_2012}. The RMS levels agree well with DNS further away from the wall, showing that the current methodology has correctly captured the modulation of near-wall turbulence by outer-layer motions that is seen experimentally \cite{HutchinsMarusic2007}. For comparison, a separate QDNS was run with only the mean mass flow and velocity gradients imposed, giving an unmodulated result (shown on figure \ref{fig:rms_uplus} with the chain dotted line) for comparison. It can be seen that the effect of modulation of near-wall turbulence by outer-layer structures is to increase the RMS levels by a factor of $\sim$2.5 at this Reynolds number. The effect of increasing RMS with Reynolds number would only be properly obtained in conventional LES using the wall-resolved approach, which would however be significantly more expensive than the current method. A wall-resolved LES grid to do the same calculation as shown here (allowing for a factor of four under-resolution in all directions compared to the reference DNS) would need 71 million grid points, compared to less than half a million employed here. The nested LES approach of \cite{Tang2013} also gives the modulation effect, but not the multi-block model of \cite{Pascarelli2000}, which uses the same replicated near-wall block everywhere on the wall.

\begin{figure}[t]
  \centering
  \includegraphics[width=0.6\textwidth]{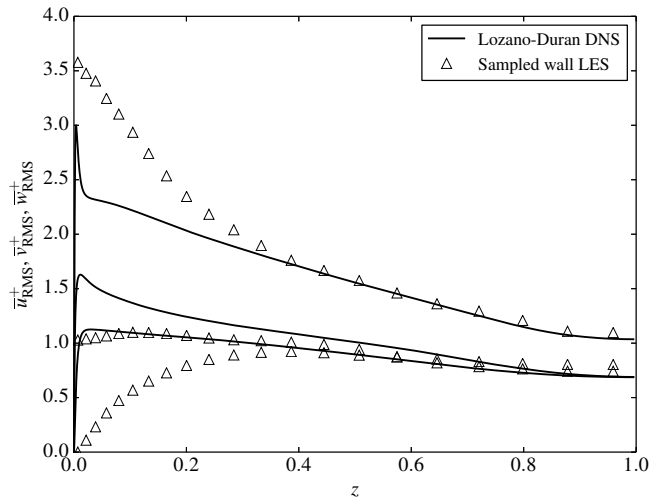}
  \caption{Root-mean-square turbulence statistics from the LES part of the simulation, compared to DNS \cite{LozanoDuran2014} at $\textrm{Re}_\tau=4200$}
  \label{fig:uvwplus_rms}
\end{figure}

The extremely coarse-grid LES shows significant errors in the structure of the turbulence as the wall is approached. Figure \ref{fig:uvwplus_rms} shows RMS values of all velocity components compared to DNS. Here, only the resolved part of the LES is shown, but nevertheless there is a significant overshoot relative to the DNS. In particular the streamwise velocity fluctuations are significantly higher and the wall-normal velocity fluctuations are significantly lower than the DNS. In both cases the effect of the wall extends to much higher values of $z$ than it should, due no doubt to the severe under-resolution of turbulence near the wall, with only larger structures resolved on the LES grid. It should be noted that the sub-grid model used here is the classical Smagorinsky model and no effort has been made to optimise the model formulation in the near-wall region. Other formulations such as dynamic Smagorinsky or WALE would be expected to do better, but the grid is so coarse in these cases that good agreement is not to be expected. A more limited expectation is that the LES resolve sufficient features of the turbulence to provide a reasonable model of the outer-flow, with the shear stress at the wall provided by the QDNS and not so dependent on the subgrid modelling (since only the local flow derivatives are passed to the QDNS as boundary conditions).

\begin{figure}[t]
  \centering
  \includegraphics[width=1.0\textwidth]{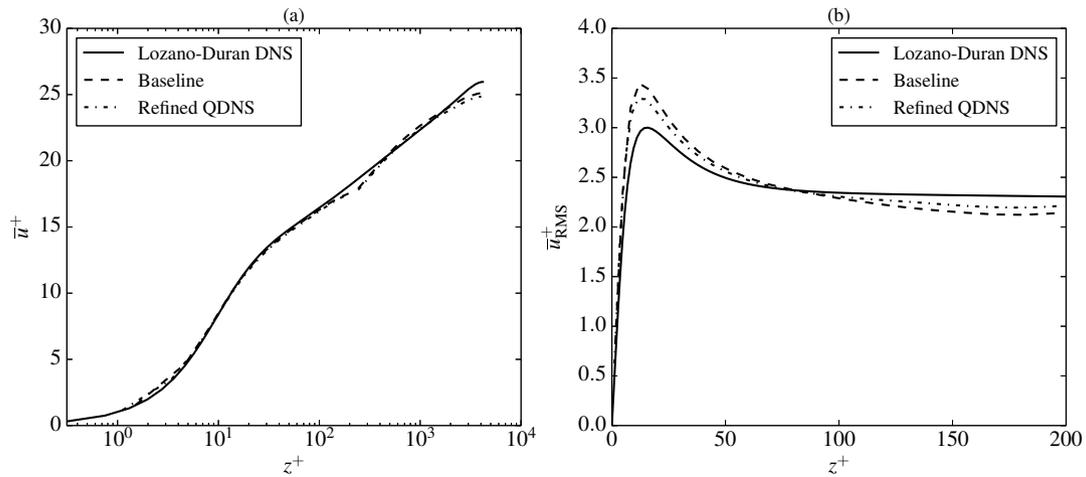}
  \caption{Sensitivity of the mean streamwise velocity and near-wall RMS streamwise velocity at $\textrm{Re}_\tau=4200$ to grid resolution of the QDNS, comparing the baseline case ($24^3$) with a refined case ($32^3$).}
  \label{fig:sensitivity_refined_qdns}
\end{figure}

\begin{figure}[t]
  \centering
  \includegraphics[width=1.0\textwidth]{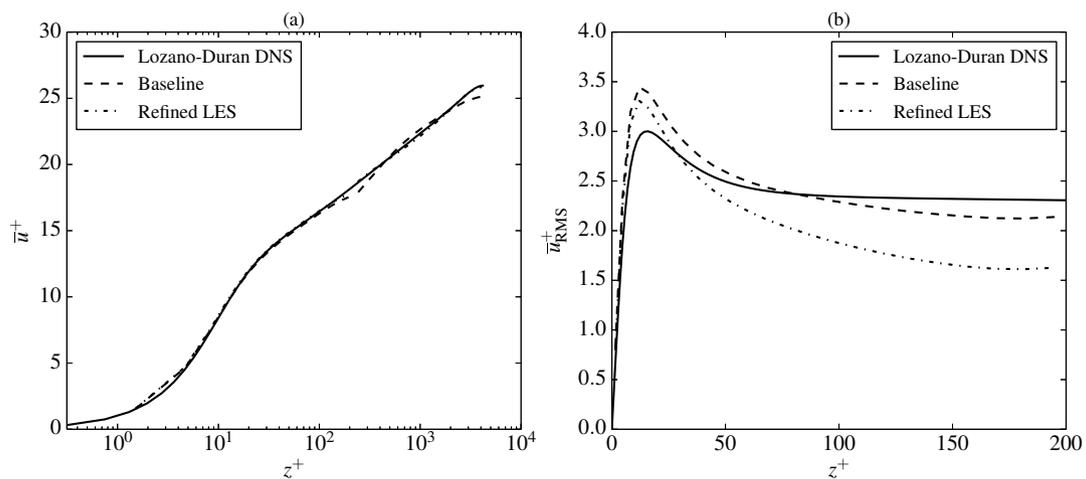}
  \caption{Sensitivity of the mean streamwise velocity and near-wall RMS streamwise velocity at $\textrm{Re}_\tau=4200$ to grid resolution of the LES, comparing the baseline case ($24^2\times 42$) with a refined case ($96^2\times 56$).}
  \label{fig:sensitivity_refined_les}
\end{figure}

\begin{figure}[t]
  \centering
  \includegraphics[width=1.0\textwidth]{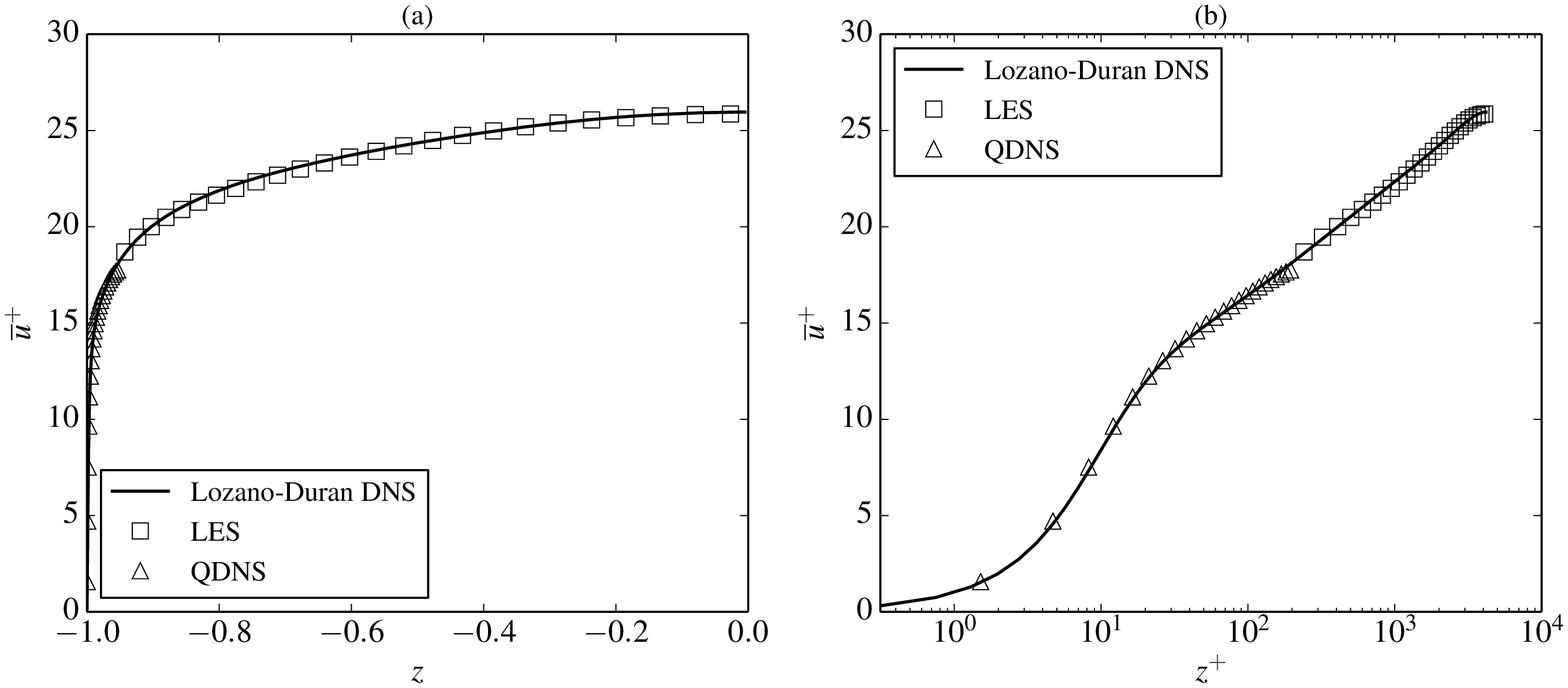}
  \caption{Comparison of the combined LES/QDNS results for mean streamwise velocity with DNS \cite{LozanoDuran2014} (solid line) at $\Rey_\tau=4200$ (a) in linear scale, and (b) in semi-logarithmic co-ordinates. Open triangles show the QDNS (carried out on $24^3$ grids), squares the refined LES (on a $96\times96\times 56$ grid).}
  \label{fig:mean_uplus_combined_refined_les}
\end{figure}

\begin{figure}[t]
  \centering
  \includegraphics[width=1.0\textwidth]{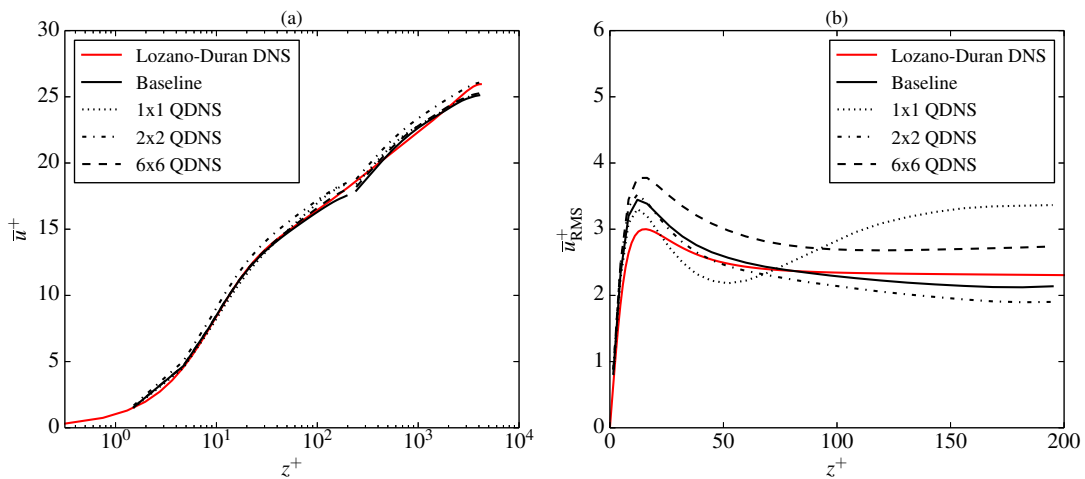}
  \caption{Sensitivity of the mean streamwise velocity and near-wall RMS streamwise velocity at $\textrm{Re}_\tau=4200$ to the QDNS arrangement, comparing the baseline case ($4^2$ on each wall) with two coarser cases ($2^2$ and $1^2$ on each wall) and one finer case ($6^2$ on each wall).}
  \label{fig:sensitivity_blocks}
\end{figure}

\begin{figure}[t]
  \centering
  \includegraphics[width=0.7\textwidth]{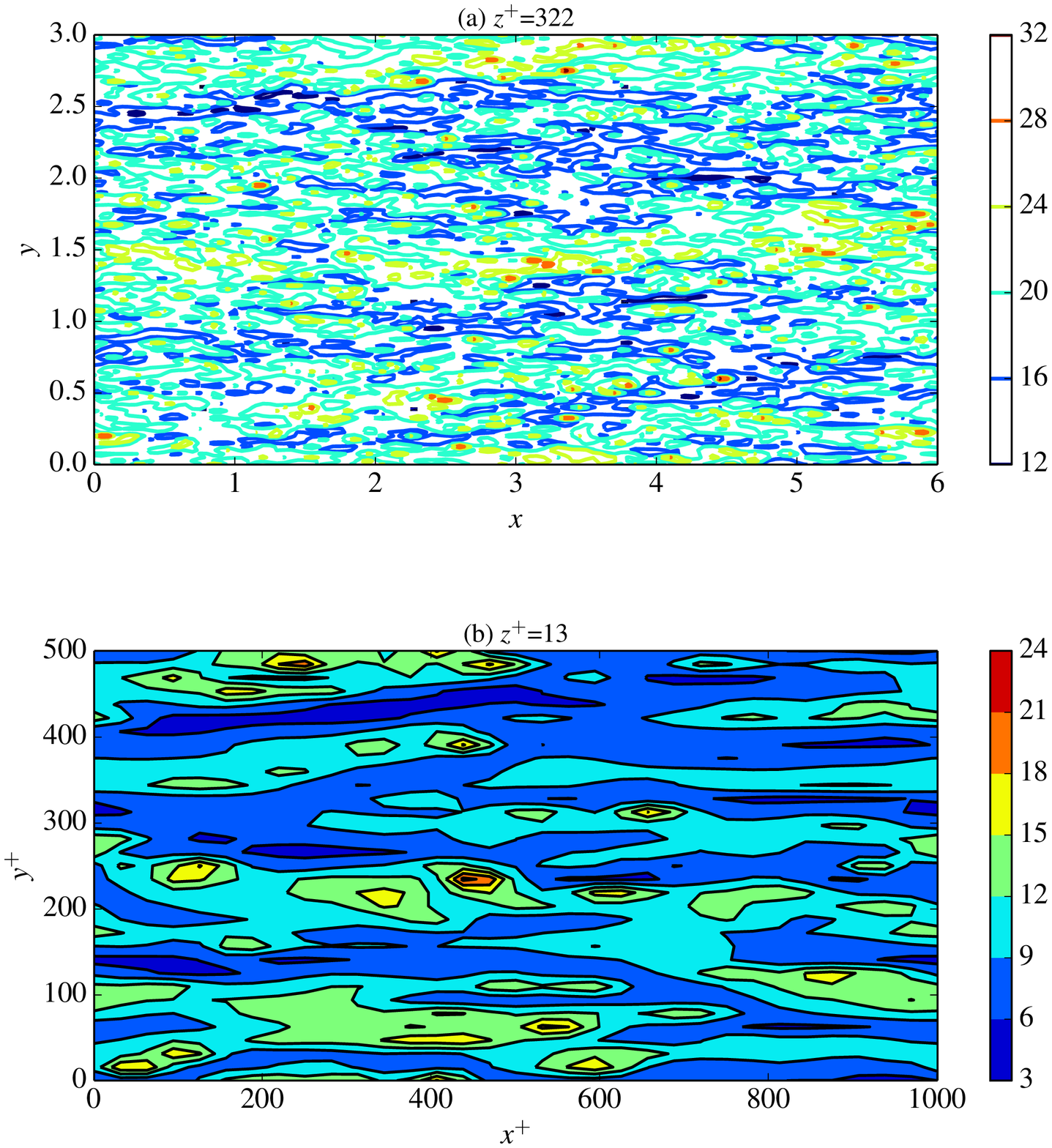}
  \caption{Plan view showing (a) streamwise velocity contour lines at $z^+=322$ from the LES at $\textrm{Re}_\tau=20000$ with the dark areas showing the locations of the QDNS domains, (b) expanded view of filled contours of streamwise velocity at $z^+=13$ in one of the QDNS sub-domains.}
  \label{fig:uplus_contours_Re20000}
\end{figure}

\begin{figure}[t]
  \centering
  \includegraphics[width=1.0\textwidth]{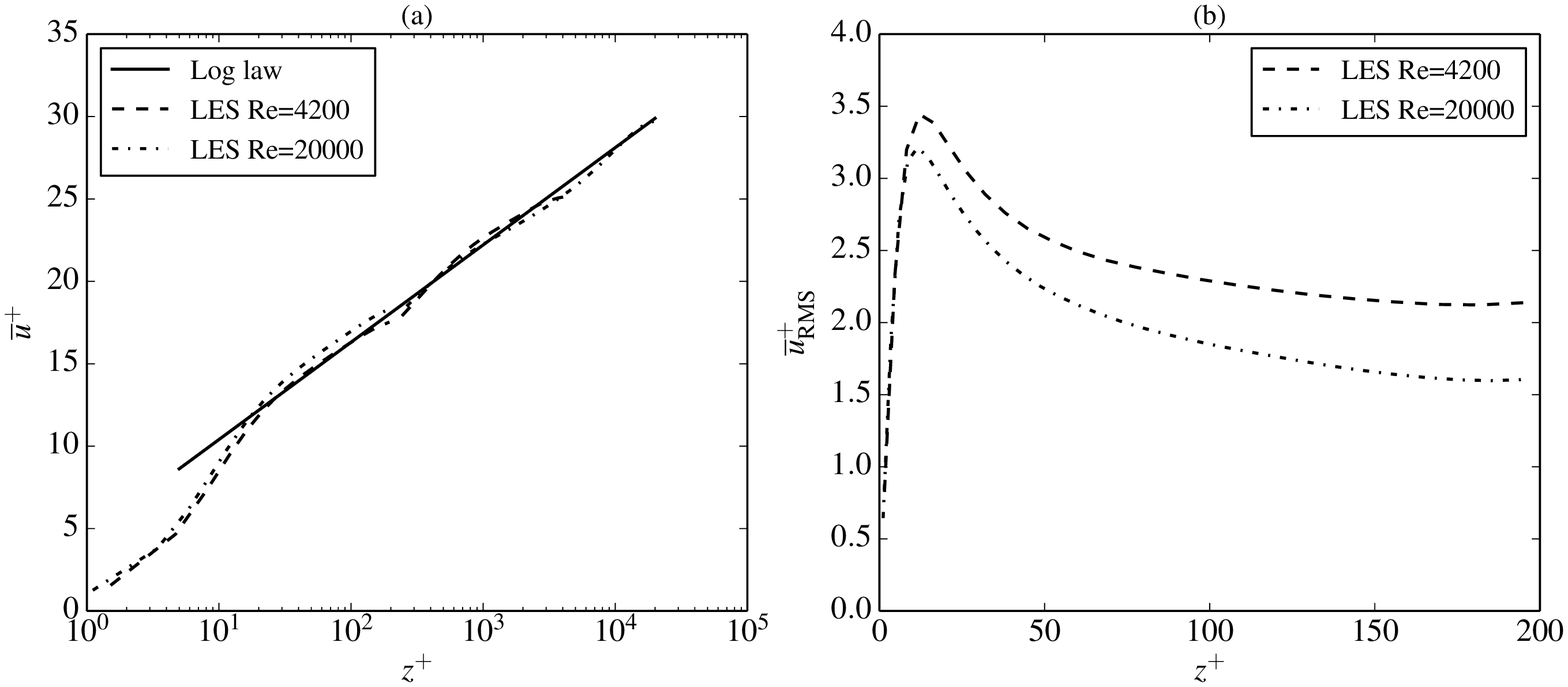}
  \caption{Comparison of (a) the mean and (b) the near-wall RMS streamwise velocities for cases at $\textrm{Re}_\tau=4200$ and at $\textrm{Re}_\tau=20000$}
  \label{fig:Re4200_Re20000_comparison}
\end{figure}

Any simulation-based model of turbulence is only useful if it provides a suitable degree of grid independency. In the current case the resolution required for the QDNS is reasonably well known, based on previous DNS. Figure \ref{fig:sensitivity_refined_qdns} (a) shows a negligible effect on the mean flow of increasing the near-wall QDNS from $24^3$ to $32^3$, which is still well below the levels required for a resolved DNS ($64^3$ would give a resolution of $\Delta x^+=15.6$, $\Delta y^+=7.8$ and a first grid point at $z^+<1$). One effect of the increased resolution is the reduced near-wall peak of the RMS streamwise velocity, shown on figure \ref{fig:sensitivity_refined_qdns}(b), with the correct trend to agree with the DNS in the limit of very fine resolution. Additionally there is a slight improvement ($<$4\%) in the RMS from around $z^+=80$ onwards.

The effect of the grid resolution of the LES in all directions is tested in figure \ref{fig:sensitivity_refined_les}(a), where the LES grid is changed from $24 \times 24 \times 42$ to $96 \times 96 \times 56$ and the stretching parameter $a$ is decreased from 1.577 to 1.28 (in order for the LES to overlap the QDNS blocks by three cells as before). This increases the LES grid point count by a factor of 21 and the timestep is reduced by a factor of 4 due to Courant number restrictions, but relatively little change is seen in the mean flow. The main effect is for the centreline velocity prediction to change from a 3\% undershoot to a $<$1\% undershoot. Similarly, the disagreement at the interface between the LES and QDNS blocks is reduced from about 5\% to 2\%. While the agreement at the near-wall peak in the streamwise velocity RMS results, shown in figure \ref{fig:sensitivity_refined_les}(b), is better when a refined LES grid is used, the same cannot be said for the results for $z^+ > \sim$25 which deviate away from the DNS data. Both RMS velocity curves from our simulations followed a trend similar to that of the DNS results (namely the initial peak in the near-wall region followed by a relatively gradual decrease further away from the wall). These RMS curves were found to be sensitive to the method of averaging the bulk velocity and velocity derivatives from the LES to enforce the mass flow rate in the QDNSs. For each block, a number of LES grid points were used for the averaging. It was observed that too small an averaging window caused the RMS velocity curve to be significantly higher than the DNS results, which was likely caused by small grid-to-grid point oscillations (in turn caused by under-resolution of the turbulence) being picked up near the wall. On the other hand, too large an averaging window can introduce turbulence smoothing, reducing the turbulent kinetic energy levels in the QDNS and therefore causing the curve to be lower than that of the DNS. The latter may have had an effect here since the number of grid points used in each averaging window ($N_x/4 \times N_y/4$) was obviously greater in the refined case (with the length and width of the averaging window remaining the same). Note that the choice of averaging size did not significantly alter the mean streamwise velocity results which were consistently better than the results from the coarser LES grid. The ultimate convergence of the LES back to the DNS would require much finer grids and large parallel simulations, which is beyond the scope of the current investigation. Nevertheless, the limited sensitivity to the grid at these very low resolutions is promising.

Finally in this section, we consider the effect of the basic arrangement of the QDNS blocks. The baseline configuration has $4 \times 4$ blocks, as sketched in figure \ref{fig:channel}. This configuration seems to be capable of resolving near-wall flow features, as illustrated by the velocity contours that were shown on figure \ref{fig:uplus_contours_Re4200}. Figure \ref{fig:sensitivity_blocks} shows the effect of reducing the number of near-wall blocks to $2 \times 2$ and $1 \times 1$, which clearly under-samples the flow features. The mean flow on figure \ref{fig:sensitivity_blocks}(a) shows that the principal effect of reducing the near-wall block count is to slightly diminish the accuracy of the near-wall turbulence. This is possibly due to aliasing effects when trying to sample the very high-frequency turbulent structures. Whilst this result is not catastrophic, it does lead to the conclusion that $4 \times 4$ blocks is probably a minimum number of blocks for a reasonable prediction of the mean flow for the current domain size. On the other hand, increasing the number of near-wall blocks to $6 \times 6$ yields an improved mean flow prediction particularly near the LES-QDNS interface. While the RMS curve for the $6 \times 6$ case displays the correct shape, the values continue to overshoot the DNS data near the wall. As already noted, these RMS values are sensitive to the averaging procedure used to enforce the mass flow rate in the QDNS blocks.

\section{Extension to higher Reynolds number}\label{sect:higher_Re}
Since the method has been proposed here as a way of simulating high Reynolds number flows, it is of interest to test the approach at even high Reynolds numbers. In this section we consider a simulation at $\Rey_\tau=20\,000$, which is a factor of nearly 5 higher than that used in the previous section. If we keep the same near-wall QDNS configuration, with $4 \times 4$ blocks, each of $32^3$ points on the same domains in wall units, we end up with sub-domains that are 0.05 long, 0.025 in the spanwise direction, with $z^+=200$ reached at $z=0.01$. Maintaining the same link between the LES and QDNS (i.e.~one $\Delta x_\textrm{LES}$ matching to the entire QDNS subdomain) as in the previous section, and retaining approximately the same stretching property of the grid (i.e.~maximum to minimum $\Delta z$) we end up with an LES grid of $120 \times 120 \times 90$. Courant number considerations again lead to a choice of 25 iterations of the QDNS per LES step, with $\Delta t_{\textrm LES}=0.00035$. Even at the higher $\Rey_\tau$ most of the cost ($>90\%$) resides in the QDNS simulations and most of the additional cost is due to the increased number of time steps required at the higher $\Rey_\tau$, which (if it works) represents a linear scaling of the total simulation cost with $\Rey_\tau$ in the channel flow example here. 

Figure \ref{fig:uplus_contours_Re20000} shows a plan view of the simulation at $\Rey_\tau=20\,000$, for comparison with figure \ref{fig:uplus_contours_Re4200} which showed the equivalent figure at $\Rey_\tau=4200$. Part (a) of figure \ref{fig:uplus_contours_Re20000} shows the streamwise velocity field from the LES at $z^+=322$, with the QDNS block superimposed, although these are too small to be clearly visible. Figure \ref{fig:uplus_contours_Re20000}(b) shows the flow in one of the QDNS blocks at $z^+=13$, showing qualitatively the same near-wall streak structure as was seen in the lower Reynolds number case. Compared with figure \ref{fig:uplus_contours_Re4200}(a), figure \ref{fig:uplus_contours_Re20000}(a) shows a much wider range of scales. The imprint of very large structures can be seen in figure \ref{fig:uplus_contours_Re20000}(a) as streamwise-elongated zones of higher- or lower-than-average streamwise velocity. Superimposed on this are smaller-scale structures down to the grid scale. On the one hand this increase in the range of scales is a more accurate picture of a turbulent flow than the picture shown in figure \ref{fig:uplus_contours_Re4200}(a), since a wider range of the turbulent energy cascade is captured. On the other hand, this picture also illustrates a possible weakness of the current approach, since the linear interpolation method used to feedback the shear stress from the QDNS to the LES will clearly not be accurate, apart from very close to the QDNS locations. 

Statistical results for the simulation at $\Rey_\tau=20\,000$ are shown on figure \ref{fig:Re4200_Re20000_comparison}, comparing the results with the logarithmic law of the wall $u^+=1/\kappa \log z^+ +b$ with $\kappa=0.39$ and $b=4.5$ (where these values have been chosen to agree with the DNS data from \cite{LozanoDuran2014}). The solution overshoots the log law by about 6\% near the LES-QDNS interface. It seems unlikely that sub-grid models can be blamed for the overshoot, although this is something that could be tested. Compared to \cite{Dean1978} the mean flow prediction is approximately 9\% too low, although we should note that the Reynolds number in this simulation is well above the highest Reynolds number used by Dean to make his correlations. In general the RMS streamwise velocity fluctuations shown on figure \ref{fig:Re4200_Re20000_comparison}(b) follow the expected trend, with the RMS increasing as $\Rey_\tau$ increases. The near-wall peak clearly increases with the fivefold increase in $\Rey_\tau$, although the RMS at $z^+=100$ decreases by 20\% before once again rising slightly above the $\Rey_\tau$ = 4200 line. Therefore, the method proposed here clearly has limitations dependent on $\Rey_\tau$, which could be mitigated through the use of finer LES resolution or more QDNS blocks which have been shown to improve the results in the $\Rey_\tau=4\,200$ case.

In summary, the effect of increasing Reynolds number is partly captured by the method presented in this section, which is based on a computational cost (including the number of time steps) that scales approximately proportional to $\Rey_\tau$. However the results at $\Rey_\tau=20,000$ deviate from the logarithmic law, suggesting that the quality of the results will decrease with further increases in $\Rey_\tau$. To improve on this probably requires more computational resource, and in this respect it is interesting that the trend to over-predict the logarithmic law of the wall was also seen when the number of QDNS blocks was reduced to $2 \times 2$ and $1 \times 1$, as shown in figure \ref{fig:sensitivity_blocks}. This suggests that one method to increase the accuracy of the simulations is to increase the number of QDNS blocks, for example to $32 \times 32$ at $\Rey_\tau=20\,000$. Although this parallelises trivially, it formally represents a scaling of the computational grid with $\Rey_\tau^2$ for channel flow, albeit with a much lower constant of proportionality than wall-resolved LES. Another method of increasing the accuracy at higher $\Rey_\tau$ would be to increase the domain size of the QDNS, also resulting in a higher scaling exponent. These estimates may be reduced if the resolution of structures associated with the mixed scaling of \cite{Moarref2013} is the limiting factor. Otherwise, for very high $\Rey_\tau$ one may need to apply the method recursively, with successively smaller domains as the wall is approached.

\section{Conclusions}\label{sect:conclusions}
A new approach to simulating near wall flows at high Reynolds number has been presented and tested. The method relies on LES for the whole domain, but with the skin friction supplied from a set of quasi-DNS of the near-wall region (out to a wall normal distance of $z^+=200$). These near-wall simulations use periodic boundary conditions and are not space-filling, but provide an estimate of the two components of skin friction, given the instantaneous near-wall velocity gradients. The method has an extremely small communication overhead between the LES and quasi-DNS and is thus suitable for scaling to large core counts. The accuracy of the method was demonstrated for a turbulent channel flow at $\Rey_\tau=4200$, for which less than half a million points were used, compared to the reference DNS that used over 4 billion points. Besides the low cost, a particular feature of the new simulation approach is that it is able to predict the effect of modulation of small-scale near-wall features by large structures, residing either in the logarithmic or outer regions of the flow. This makes it possible, for example, to study the effects of wall-based flow control schemes in a high-Reynolds number external environment. The method is found to be robust to changes in grid resolution. An O($\Rey_\tau$) total cost extrapolation to $\Rey_\tau=20\,000$ demonstrated some limitations, suggesting that accurate simulations at higher $\Rey_\tau$ probably have a higher total cost scaling (including an increase in grid points and in the number of timesteps), however at much lower cost relative to wall-resolved LES. For the particular case considered here, that of turbulent channel flow, wall functions for LES based on the logarithmic law of the wall would be expected to work well. The advantage of the current approach is that the log law is not assumed and it would be expected that the effects of a range of non-equilibrium flow conditions could be captured, so long as the surface sampling is sufficient relative to the dominant large-scale structure in the flow. Overall the new method offers the potential for engineering calculations at high Reynolds number at a substantially lower computational cost compared to current LES techniques.

\ack A significant part of this work was first presented at the 8th International Conference on Computational Fluid Dynamics (ICCFD); see \cite{SandhamJohnstone2014}. CTJ was supported by a European Commission Horizon 2020 project grant entitled ``ExaFLOW: Enabling Exascale Fluid Dynamics Simulations'' (grant reference 671571). RJ was supported partially by the UK Turbulence Consortium (EPSRC grant EP/L000261/1). The authors would like to acknowledge the support of iSolutions at the University of Southampton and the use of the in-house Iridis 4 compute cluster. The data files generated as part of this work will be available through the University of Southampton's institutional repository service.

\bibliography{heterogeneous-modelling-les-qdns}
\bibliographystyle{wileyj.bst}
\end{document}